GRAPHENE RADIO: DETECTING RADIOWAVES WITH A SINGLE ATOM SHEET

M. Dragoman[1], D. Neculoiu[2], A. Cismaru[1], G. Deligeorgis [3], G. Konstantinidis[4], D. Dragoman[5*]

[1] National Institute for Research and Development in Microtechnology (IMT), P.O. Box 38-160, 023573 Bucharest, Romania,

[2] Politehnica University of Bucharest, Electronics Dept., 1-3 Iuliu Maniu Av., 061071 Bucharest, Romania,

[3] LAAS CNRS, 7 Avenue du Colonel Roche, 31077 Toulouse Cedex 4, France,

[4] Foundation for Research & Technology Hellas (FORTH) P.O. BOX 1527,Vassilika Vouton, Heraklion 711 10, Crete, Greece,

[5] Univ. Bucharest, Physics Dept., P.O. Box MG-11, 077125 Bucharest, Romania.

**Abstract**

We present the experimental evidence of RF demodulation by a graphene monolayer embedded in a coplanar structure. The demodulator was tested in the frequency range from 100 MHz to 25 GHz using amplitude modulated input signals. An input power of 0 dBm (1 mW) was used which is the typical power emitted for short range wireless communication systems, such as Bluetooth. The graphene demodulator exhibits good signal response in the frequency range associated to industrial, scientific and medical (ISM) radio band (2.4 GHz).

_______________________________________________________________

*Corresponding author: danieladragoman@yahoo.com

The radio is moving ahead towards nanoscale devices and is termed as nanoradio. Its extremely miniature dimensions pave the way for new applications of nanoradio are opened not only in free-space (Bluetooth devices or RFID tags) but in blood stream for controlled drug delivery , implant monitoring[1,2] and energy harvesting [3].

Up to now, the nanoradio was based on NEMS (nano-electromechanical systems) based on carbon nanotubes which are 4-5 orders of magnitude smaller than the actual radios implemented with advanced Si technologies. The carbon nanotubes (CNTs) are cantilevers which were electrostically actuated and had resonance frequencies in the range 100 MHz-5 GHz. The first nanoradio[4] was based on such a NEMS resonator which has performed all the functions of a radio i.e. the carrier signal generation and modulation and detection. The CNT emitter and the cathode were sealed in a high vacuum enclosure. This nanoradio had serious drawbacks: the bias was around 200 V and the tunning was only 4 MHz. On the contrary, the tunneling nanoradio [5] had a larger tunability covering the FM or AM radio bands and could be biased with batteries, but its fabrication is very difficult. The nonlinear current−voltage dependence of CNT was used to demodulate the RF signal, but the demodulated current was very low (few nA) while parasitic capacitances hindered operation beyond 2 GHz[6]. A double–clamped CNT in FET configuration was recently proposed for the demodulation of RF signals[7]. AM and FM demodulation were demonstrated, but the detected current was very weak 0.1-0.2 nA.In fact, a lock-in IF amplifier was used to prove the concept.

Although the CNT-based NEMS have mechanical resonant frequencies within the RF spectrum (0.05-5 GHz) and high quality factors ($10^3$) the  resulting NEMS based radios demonstrated  lack of robustness, narrow bandwidths and weak output currents, all of which combined lead to serious noise issues. Therefore, it is uncertain if these devices will be ever used in real-life RF systems in which external mechanical vibration and external RF noise are always present. Moreover, there are numerous Si based electronic devices and integrated circuits that can be readily fabricated while exhibiting higher performances. So, the challenge is to find other nanomaterials able to maintain the small size of NEMS radios, but capable of working in larger bandwidths, and providing higher amplitudes of the detected signals.

Thus, along this line, this paper reports on the investigation of the capability of a graphene monolayer flake to detect radio waves in the RF spectrum and beyond. The graphene radio wave detector was fabricated on a high-resistivity Si substrate with resistivity greater than 8 k$\Omega$, on which 300 nm of SiO2 were grown by thermal oxidation. The graphene monolayer deposition on Si/SiO$_2$ substrate and the structural characterization (optical images, Raman spectroscopy) were performed by Graphene Industries. Over the graphene monolayer three parallel metallic electrodes forming a coplanar waveguide (CPW) were patterned. RF pads were connected to both ends of the CPW to facilitate easy connection with probe station. In the CPW configuration, the central conductor is the signal electrode and the outer conductors are ground electrodes. The first port of the device was used to inject the microwave signal in the graphene based detector and the second port was used to collect the detected signal. This structure may in fact be directly integrated with pre-existing Si devices and circuits. The detection of the modulated microwave signal is based on the nonlinear current-voltage dependence of the graphene when connected with gold contacts. This nonlinear I-V dependence has the shape of two leaky back-to-back Schottky diodes as is described in Ref. 8 and the references herein. Fabrication details are fond in Ref.8.

The graphene radio wave detector is presented in Fig. 1 and its DC current-voltage (I-V) characteristic are shown in Fig. 2. The observed nonlinear  I-V dependence is modeled as two leaky back-to-back Schottky diodes (see Ref.8 and the references herein).This inherent nonlinearity will be further exploited in this paper for radio wave detection. The set-up for detecting radio-waves using a graphene loaded CPW presented in Fig. 1 follows the configuration of a radio receiver. The graphene radio wave detector was excited by an Agilent E8257D with an AM modulated carrier signal.  The carrier frequency signal was varied in the frequency range 100 MHz-50 GHz and AM modulation signal was within audio signals frequency. The graphene radio wave detector was biased by a current source via a bias tee. The graphene radio wave is placed on the on-wafer probe station and its input and output are connected to the AM signal generator and the low noise amplifier (LNA) , respectively by probe tips. The detection signal was monitored by the LNA(Stanford Research SR560) and a digital oscilloscope (Tektronix , SR560) able to record the data.

We have used in the experiment carrier radio waves frequencies in the frequency range of 100 MHz to 25 GHz at a constant power of 0dBm (1mW) which corresponds to the emitted power of Bluetooth standard (Class 3) radio, 1-5 m range. The above frequency range encompasses a huge frequency range i.e. RF spectrum (starting with VHF) and the microwave spectrum (L-K bands). The modulation is 1KHz which characterizes the human voice and is also the central frequency of many audio hearing aids devices. The radio-wave detected voltage dependence on the frequency and current bias is displayed in Fig. 3 at various bias currents. We can see that the maximum values of detected voltages are located in the range 1-5 GHz where the majority of wireless applications are working today.Even without current bias, the graphene radio wave detectors exhibits a detected voltage in the range 0.005-0.2 V in the range 3-10 GHz spectrum (see Fig. 4). In Fig. 5, we have displayed the detected signal in time of AM radio wave signal. Due to the limitations of the measurement setup higher modulation frequency was not measured but indications exists that it could be detected as well with similar results. In conclusion, AM modulated radio waves were demodulated at room temperature in the frequency range 0.1-25 GHz by a simple detector device based on a monolayer graphene embedded in a coplanar line. This concept paves the way towards simplified nanodetctors for both digital as well as analog electronics while retaining CMOS compatibility.

*Acknowledgements* M.Dragoman states that this work was supported by a grant of Romanian National Authority for Scientific Research, CNCS-UEFISCDI, project number PN-II-ID-PCE-2011-3-0071. G. Konstantinidis acknowledges the "Graphene Center" of FORTH while all the authors the LEA SMARTMEMS.

Figure captions

Fig. 1 The graphene radio waves detector.

Fig. 2 The graphene coplanar detector current-voltage dependence

Fig. 3 The detected DC voltage as a function of frequency for various biases :1 mA (black) ; 2

mA (blue), 3 mA (red).

Fig. 4 The detected DC voltage as a function of frequency at zero DC bias.

Fig. 5 The detected DC voltage evolution in time.

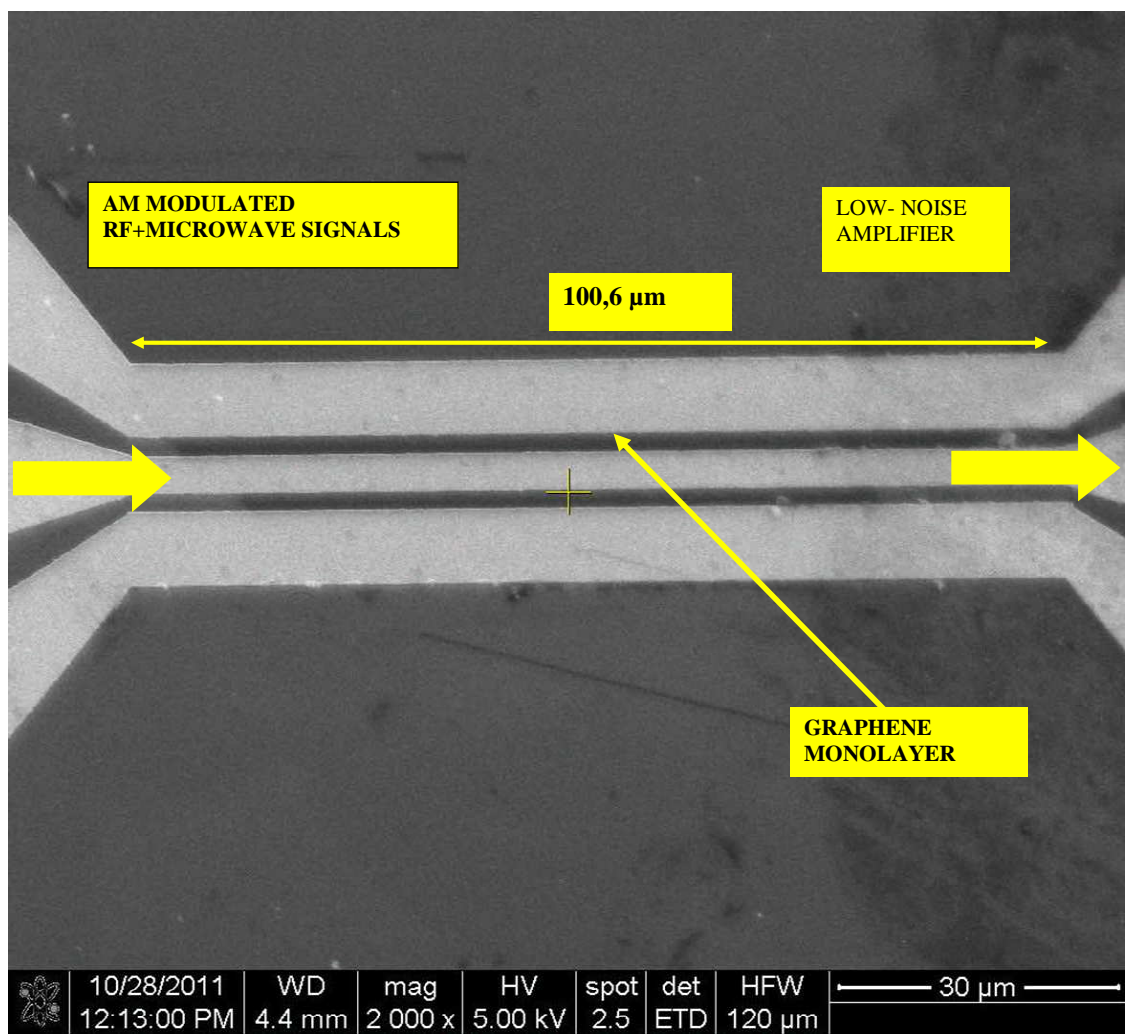

Fig. 1 DRAGOMAN

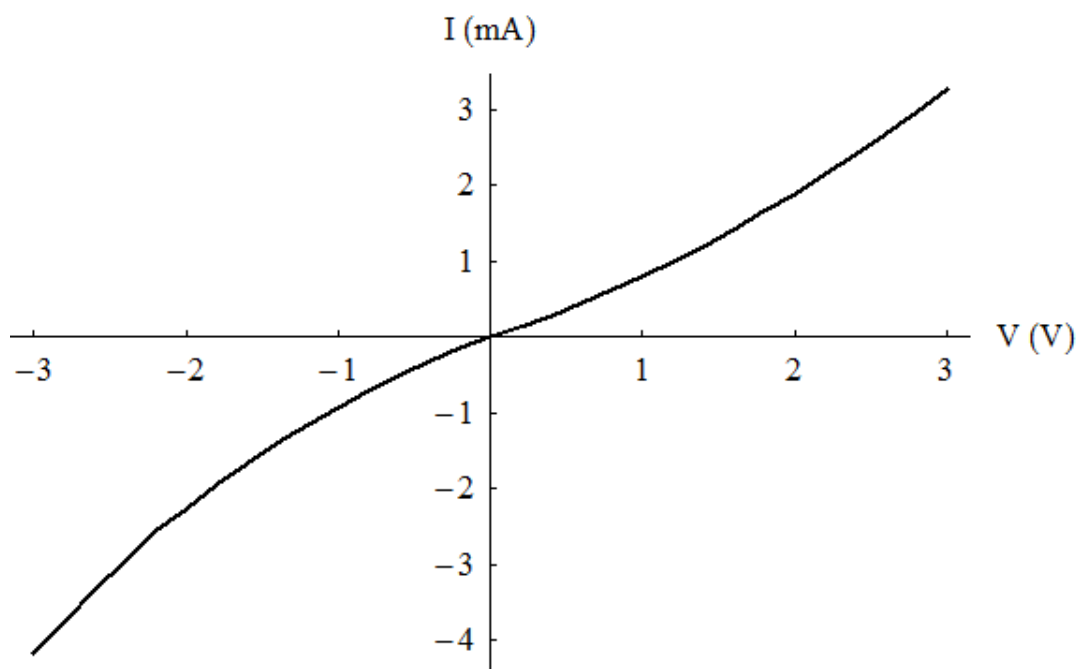

Fig. 2 DRAGOMAN

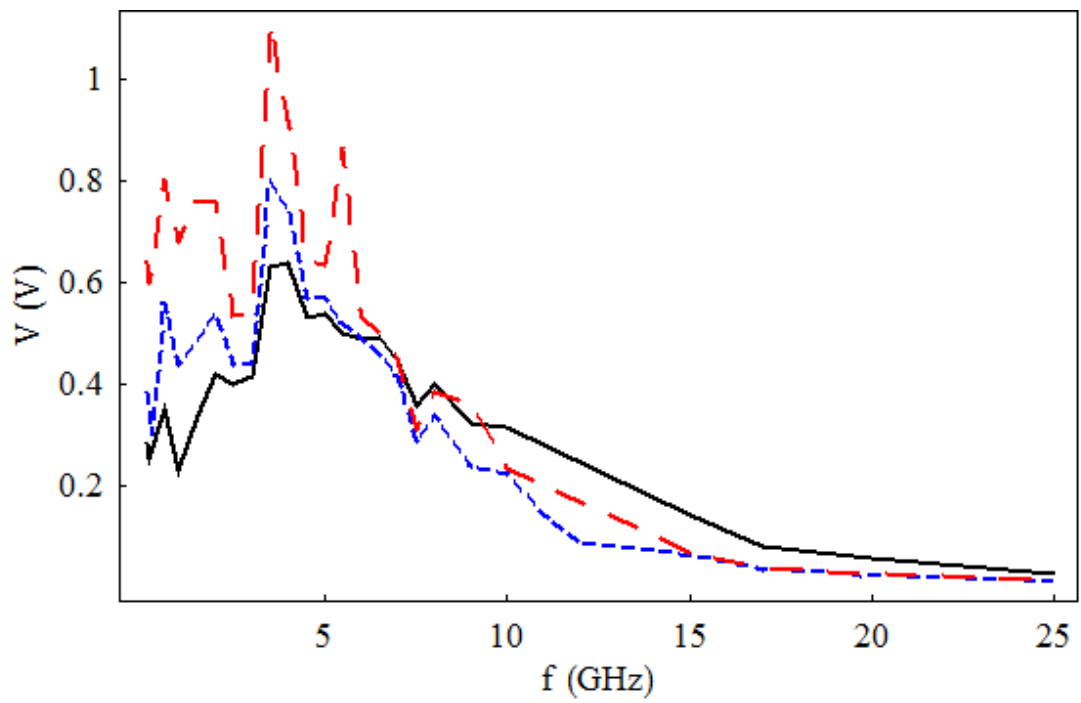

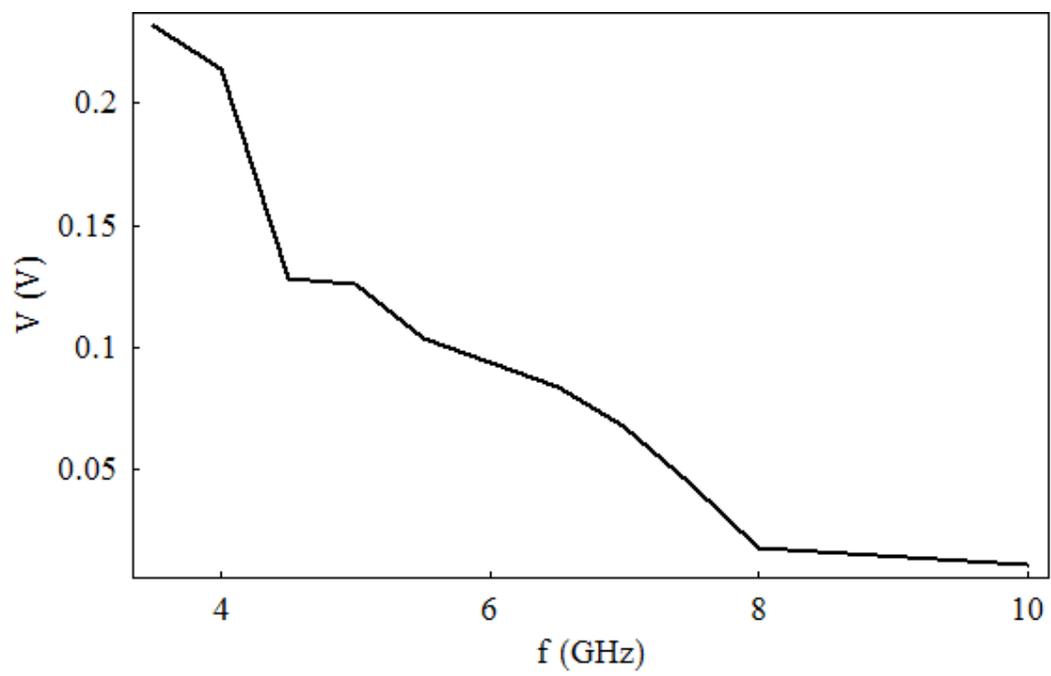

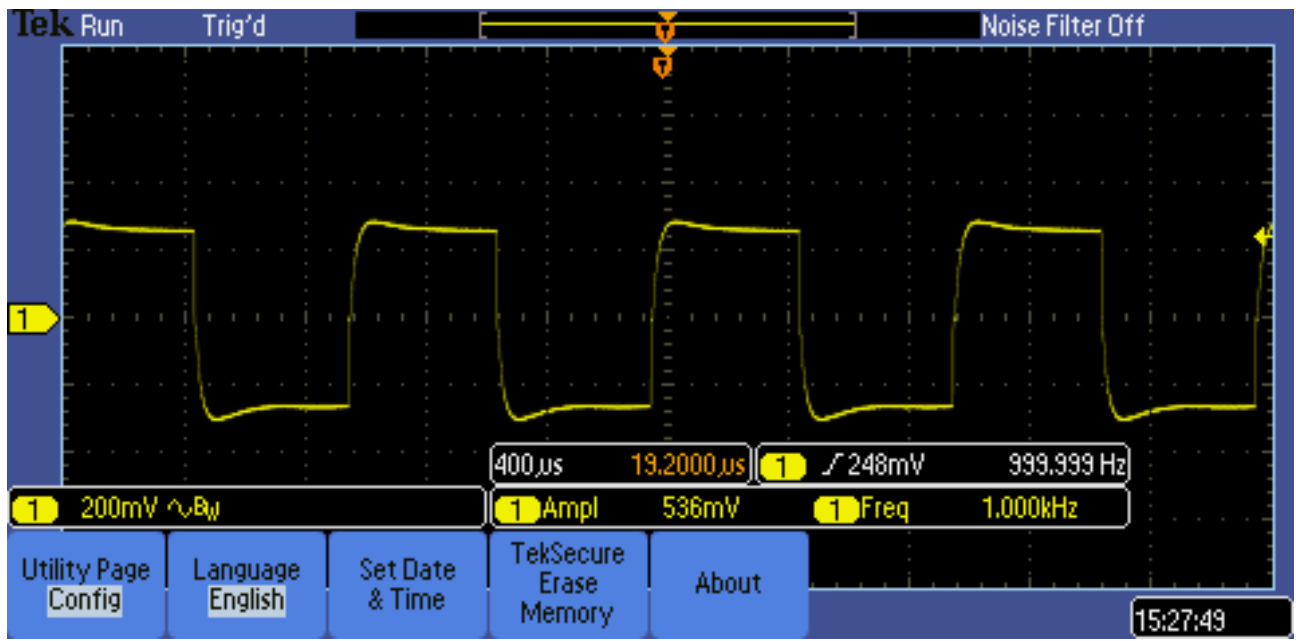

Fig. 5 DRAGOMAN